\documentclass[twocolumn,aps,prx]{revtex4}

\usepackage{graphics}
\usepackage{graphicx}
\usepackage{dcolumn}
\usepackage{bm}
\usepackage{amsmath,amssymb,epstopdf}
\usepackage{color}

\newcommand{\tr}[1]{\textrm{Tr}\left( #1 \right)}

\newcommand{\de}{\partial}

\newcommand{\eq}[2]{\begin{equation} \label{#1} #2 \end{equation}}
\newcommand{\eps}{\epsilon}

\newcommand{\etal}{{\em et al.}}

\newcommand{\vvv}{\mathbf{v}}

\newcommand{\schr}{Schr\"odinger }
\newcommand{\schw}{Schwarzschild }
\newcommand{\rn}{Reissner-Nordstr\"om }
\newcommand{\R} {\text{Re}}

\begin{document}

\title{Universal quantum Hawking evaporation of integrable two-dimensional solitons}
\author{Charles W. Robson, Leone Di Mauro Villari and Fabio Biancalana}
\affiliation{School of Engineering and Physical Sciences, Heriot-Watt University, EH14 4AS Edinburgh, UK}
\date{\today}

\begin{abstract}
We show that any soliton solution of an arbitrary two-dimensional integrable equation has the potential to eventually evaporate and emit the exact analogue of Hawking radiation from black holes. From the AKNS matrix formulation of integrability, we show that it is possible to associate a real spacetime metric tensor which defines a curved surface, perceived by the classical and quantum fluctuations propagating on the soliton. By defining proper scalar invariants of the associated Riemannian geometry, and introducing the conformal anomaly, we are able to determine the Hawking temperatures and entropies of the fundamental solitons of the nonlinear Schr\"odinger, KdV and sine-Gordon equations. The mechanism advanced here is simple, completely universal and can be applied to all integrable equations in two dimensions, and is easily applicable to a large class of black holes of any dimensionality, opening up totally new windows on the quantum mechanics of solitons and their deep connections with black hole physics.
\end{abstract}

\maketitle

\section{Introduction}

In a now half-forgotten work published in 1976, Abdus Salam and his student John Strathdee proposed what was at that time a revolutionary connection between two different fields, namely Einstein's theory of general relativity and the theory of nonlinear evolution equations \cite{salam}. Their conjecture was simple but at the same time profound: {\em a black hole is nothing else than a soliton}. Black holes are localised, finite energy vacuum solutions of Einstein's equations of general relativity, which are very nonlinear and are written in terms of the metric and the Ricci tensors. On the other hand, it was very well known that nonlinear equations can have localised solutions called ``solitary waves", i.e. waves that do not change shape during time evolution. Certain special nonlinear evolution equations, somewhat misleadingly called {\em integrable equations}, have an infinite number of conservation laws and are in principle completely solvable analytically; their solitary waves have special stability properties: these waves are called {\em solitons}, and possess a finite number of parameters that completely determine their identity \cite{drazin}. Salam and Strathdee's conjecture therefore implies that a black hole is nothing other than one type of soliton, and as such it can be studied using soliton methods, which we briefly describe below. This conjecture is far-reaching, and effectively opened the field of {\em gravitational analogues}. This fascinating connection was further deepened by Zakharov and Belinski\^i in 1978, who showed for the first time that there are soliton solutions of the Einstein equations in 2D (typically exhibiting cylindrical symmetry, such as wormholes) that can be found by means of the inverse scattering method \cite{Belinsky}. In light of this connection the so-called ``no-hair theorem", i.e. the fact that all black hole solutions of the Einstein-Maxwell equations of gravitation and electromagnetism in general relativity can be completely characterized by only three externally observable classical parameters (mass, electric charge, and angular momentum), perfectly fits into the solitonic framework, since solitons always have a finite, typically small, number of parameters which completely defines their properties: solitons with identical parameters behave identically and have no structure, exactly as black holes.

Historically, one of the most interesting and studied aspects of black holes is their ability to radiate quantum mechanically due to conversion of energy stored in spacetime curvature into radiation, until eventually all of their energy and mass is radiated away. This outward flux of radiation can be measured at infinity with a characteristic temperature known as the Hawking temperature \cite{hawking}. Since Hawking's seminal work other thermodynamical properties of black holes have been described in detail, perhaps most significantly the entropy, which in four dimensions turns out to be proportional to the black hole surface area, an observation due to Bekenstein which predates Hawking's work \cite{bekenstein}. This result has had major implications for modern physics, and as its study requires a combination of general relativity, quantum field theory, thermodynamics and information theory, it is considered a promising route towards a deeper understanding of quantum gravity \cite{susskind}.

Much literature has been devoted, experimentally and theoretically \cite{faccio,ulf,jeff}, to the possible detection of Hawking radiation in classical or semiclassical analogue systems (typically optical, hydrodynamical or based on quantum condensates), due to the considerable (and probably insurmountable) difficulty of detecting it in a real astrophysical setting. In light of Salam and Strathdee's conjecture about the equivalence between solitons and black holes, it is suggestive to imagine that the {\em exact analogue} of Hawking radiation can be directly seen in the physics of solitons. It is however important to establish a precise, unambiguous mathematical and physical connection between the two objects, something that we shall do in the present paper for the first time.

In this paper, following Salam's steps, we push the analogy between black holes and solitons to a new level. We show that any soliton solution of a two-dimensional integrable nonlinear evolution equation potentially possesses a Hawking temperature which is determined solely by the geometrical properties of an internal surface connected to the specific soliton, called the {\em integrable surface}. The curvature of this surface is turned into Hawking radiation due to the existence of a {\em quantum anomaly}, well-known in two-dimensional quantum field theory systems (the so-called conformal field theory, or CFT). This anomaly manifests as a non-zero trace of the energy momentum tensor of quantum fields in a black hole background. Hawking radiation from the 2D black hole can be described either in terms of a conformal anomaly \cite{conformal}, an anomalous non-zero energy-momentum tensor trace of conformal matter fields in a black hole background, or more generally \cite{wilczek} can be described in terms of a gravitational anomaly, emerging from a breaking of the symmetry between the ingoing and outgoing horizon modes, again producing a non-zero energy-momentum tensor trace. In essence, the quantum fluctuations propagating on the classical soliton feel the curvature of the integrable surface, and the energy stored in this curvature is then converted into radiation modes, with a Hawking temperature dictated by the quantum anomaly.

The plan of the paper is the following. In Section \ref{sec2} we briefly introduce the well-known AKNS method for integrable nonlinear evolution equations in 2D. The AKNS matrices play a crucial role in the construction of the metric of the solitonic surface, which is defined and discussed in Section \ref{sec3}. In this section we also give a quick summary of the curvature invariants that are used in parts of our construction of the Hawking temperature for solitons. In Section \ref{sec6} we start discussing the Hashimoto metric for the important (and slightly subtle) case of the nonlinear \schr equation, applying our method to calculate the Hawking temperature of its fundamental bright soliton. In Section \ref{sec7} we show how the same procedure is applied to the KdV and sine-Gordon fundamental solitons. In Section \ref{sec8} we introduce the new concept of ``quantum soliton thermodynamics": for each example considered, we establish the thermodynamical laws of the soliton `black hole' and in this way we are able to calculate explicitly the soliton entropies and completely confirm the previous Hawking temperature calculations in an entirely different way. Section \ref{sec9} is devoted to the application of our formalism to a real-world scenario of a soliton propagating inside an optical fiber, and we give numerical estimates for the Hawking temperature and the soliton lifetime for the nonlinear \schr equation soliton along with some ideas for a possible experimental verification of the effect.

The implications of our work are that we {\em completely identify} the soliton and the black hole concepts in two-dimensions. In reality, this complete identification is extended also to a large class of arbitrarily higher-dimensional black holes. In that case, a simple dimensional reduction will lead to a two-dimensional (not necessarily diagonal) metric tensor that can be studied with our technique.

\begin{figure}
\includegraphics[width=8cm]{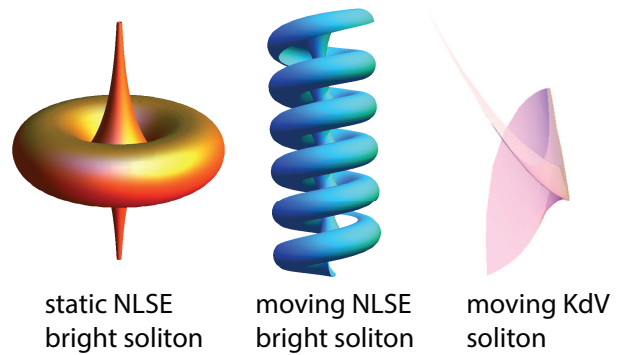}
\caption{Integrable surfaces for some important solitons of integrable equations: (a) for the static ($v=0$) bright soliton of the NLSE, showing that at $\pm\infty$ one has a Beltrami pseudosphere with constant negative curvature, while near the soliton peak we have a ``doughnut" with positive curvature; (b) for a moving ($v\neq0$) bright NLSE soliton; and (c) for the moving KdV fundamental soliton.\label{fig1}}
\end{figure}

\section{Integrability and AKNS matrices}\label{sec2}

The highly non-trivial method of the inverse scattering transform (IST) was invented in 1967 by Gardner \etal\  \cite{Gardner} in order to turn the problem of solving a {\em nonlinear} partial differential equation (PDE) into a set of {\em linear} equations for the scattering data of a quantum mechanical problem. This allowed the reconstruction of the \schr potential from the scattering data (the so-called `inverse problem') and gave a methodology to find analytical soliton solutions of nonlinear equations. This method only works for a certain class of nonlinear PDEs, namely {\em integrable equations}: the defining property of such equations is that they have an {\em infinite} number of conservation laws and solutions, and that {\em all} solutions can in principle be found analytically -- although the procedure is often so long and tedious that only the calculation of the simplest solutions is practical \cite{drazin,ablowitz}. The concept of integrability is particularly delicate in systems with infinite degrees of freedom. In particular, the existence of an infinite set of conserved quantities with vanishing Poisson brackets might not be enough; in layman's terms we could say that the infinity of the conservation laws might not match the infinity of the dimensionality. In this complicated scenario a more profound interpretation of the IST provides a suggestive solution. In 1971 Zakharov and Faddeev proved for the first time that the IST can actually be interpreted as a canonical transformation to action-angle variables \cite{Zakharov1}. This argument was further generalised later \cite{Zakharov2,Kodama} to provide a fascinating extension of the Arnold-Liouville theorem to nonlinear PDEs solvable via the IST. A few years previously, in 1968, Lax \cite{lax} showed that for each integrable equation it is possible to find two nontrivial operators, now called the {\em Lax pair}. The compatibility condition of the linear equations associated to the Lax pair defines the nonlinear evolution equation of interest \cite{ablowitz}.

In 1973, Ablowitz \etal\  generalized Lax's method to matrix operators (for a review of the idea and the key papers see Ref. \cite{AKNS}). This method, which we use in this paper, is now called the AKNS method, after the authors' initials in their seminal work. The AKNS method is nowadays used to solve the initial value problem for a remarkably large class of physically interesting nonlinear evolution equations in 2D. Below we give a quick sketch of the AKNS method.

Consider the two {\em linear} equations ($x$ and $t$ subscripts mean derivatives with respect to these variables)
\begin{eqnarray}
\vvv_{x}&=&\hat{g}_{1}\vvv \label{eq1}\\ 
\vvv_{t}&=&\hat{g_{2}}\vvv \label{eq2},
\end{eqnarray}
where $\vvv$ is an $n$-dimensional vector and $\hat{g}_{1,2}$ are two $n\times n$ matrices (the so-called AKNS matrices). Here, $t$ and $x$ are two independent variables, and in the following we shall assume that $t$ is the evolution variable of the nonlinear equation of interest.  The compatibility condition for Eqs. (\ref{eq1}-\ref{eq2}) is obtained deriving Eq. (\ref{eq1}) with respect to $t$ and Eq. (\ref{eq2}) with respect to $x$, and imposing $\vvv_{xt}=\vvv_{tx}$. One finds:
\eq{comp1}{\hat{g}_{1,t}-\hat{g}_{2,x}+[\hat{g}_{1},\hat{g}_{2}]=0.}
Equation (\ref{comp1}) is then satisfied if the nonlinear PDE of interest is also satisfied. As an example, let us show the $2\times 2$ AKNS matrices for the nonlinear \schr equation (NLSE) \cite{ablowitz}:
\begin{eqnarray}
\hat{g}_{1}&=& \left(\begin{array}{cc} i \lambda & u \\ -u^{*} & -i \lambda \end{array}\right), \label{akns1} \\
\hat{g}_{2}&=& \left(\begin{array}{cc} -2i \lambda^{2}+ i |u|^{2} & -2\lambda u+iu_{x} \\
2 u^{*} \lambda + iu^{*}_{x} & 2i \lambda^2 - i |u|^{2}\end{array}\right), \label{akns2}
\end{eqnarray}
where $u(x,t)$ is a complex function and $\lambda$ (which is assumed to be real for the moment) is the so-called {\em spectral eigenvalue}, which encodes information on the soliton parameters. Inserting these two matrices into the compatibility condition Eq. (\ref{comp1}) gives a matrix for which the diagonal elements vanish and the antidiagonal elements are proportional to the NLSE and its complex conjugate:
\eq{nlse1}{iu_{t}+u_{xx}+2|u|^{2}u=0,} where $u\equiv u(x,t)$ is now understood to be a solution of the NLSE. For the NLSE, the AKNS matrices $\hat{g}_{\mu}$ (with $\mu=\{x,t\}$) belong to the Lie algebra of the SU(2) symmetry group of complex $2\times 2$ anti-Hermitian traceless matrices, $\mathfrak{su}$(2). It is an interesting observation (albeit somewhat trivial for the reader with knowledge in field theory) that Eq. (\ref{comp1}) can be written more implicitly as $F_{\mu\nu}\equiv\de_{\mu}\hat{g}_{\nu}-\de_{\nu}\hat{g}_{\mu}+[\hat{g}_{\mu},\hat{g}_{\nu}]=0$, which is analogous to a fictitious non-abelian gauge field with potential $\hat{g}_{\mu}$ in the chosen Lie algebra [in the NLSE case $\hat{g}_{\mu}\in\mathfrak{su}$(2)], which possesses a vanishing `curvature' tensor $F_{\mu\nu}$. Therefore Eq. (\ref{comp1}) is also known in the literature as the zero-curvature condition \cite{ablowitz}.

\section{Soliton surface metric and curvature invariants} \label{sec3}

A remarkable connection between the AKNS matrices and the geometric theory of surfaces was first discovered in 1976 in the work of Lund and Regge, and of Pohlmeyer \cite{Lund,Pohlmeyer}. They established a connection between the geometry of a privileged class of surfaces, known as integrable surfaces, and soliton theory. The next major development came in 1982 with the formal introduction of the concept of a soliton surface by Sym \cite{Sym}. The surfaces associated with the sine-Gordon and NLSE equations are pseudospherical and Hashimoto surfaces respectively \cite{rogers}.

The metric tensor  associated to a soliton surface is found to be \cite{rogers}:
\eq{metric1}{g_{\mu\nu}\equiv \alpha\tr{\de_{\lambda}\hat{g}_{\mu}\de_{\lambda}\hat{g}_{\nu}},} and the line element for this metric is the usual $ds^{2}=g_{\mu\nu}dx^{\mu}dx^{\nu}$.
The constant factor $\alpha$ in front of expression (\ref{metric1}) is due to the specific normalization of the generators of the symmetry group under consideration. For the SU(2) group, if we choose the generators to be $T_{k}\equiv\sigma_{k}/(2i)$, where $\sigma_{k}$ are the three Pauli matrices, the normalization factor is $\tr{T_{i}T_{i}}=-1/2$ ($i=1,2,3$), which results in $\alpha\equiv[\tr{T_{i}T_{i}}]^{-1}=-2$.

For the few important representative examples that we shall study in this paper, following Ref. \cite{rogers} we show the Hashimoto surfaces for the stationary bright soliton of the NLSE [Fig. \ref{fig1}(a)], the moving soliton of the NLSE [Fig. \ref{fig1}(b)], and the integrable surface for the moving soliton of the KdV equation [Fig. \ref{fig1}(c)] \cite{Tek}.

How to interpret the metric (\ref{metric1}) of the soliton surface? Is it just an arbitrary mathematical surface (which is the traditional interpretation in mathematics textbooks and previous literature), or does it contain some deeper physical information on the soliton structure? One of the major results of this paper, and in our opinion a groundbreaking one, is that this metric represents {\em a real spacetime metric} perceived by the classical or quantum fluctuations propagating on the soliton. This claim is based on connecting the metric of Eq. (\ref{metric1}) with the quantum mechanical metric describing the geometry of the Hilbert space  \cite{Provost,Ruppeiner}. Reference \cite{Provost} shows cases for which the metric structure is fixed by the quantum fluctuations. In Ref. \cite{Ruppeiner} instead the metric is related to the thermal fluctuations of the system, and can be therefore applied to classical fluctuations. The comparison of these two results establishes a connection between the classical fluctuation-dissipation theorem and the metric defined by the fluctuations of the system.

We now formalise the link between the quantum mechanical metric and the classical metric on the soliton surface. Consider a general Hilbert space $\mathcal{H}$ and an arbitrary quantum state $\psi(x,t)$. The distance between two infinitesimally close quantum states belonging to $\mathcal{H}$, that is $||\psi(x + dx, t + dt)-\psi(x,t)||$, induces a metric of the form
\begin{equation} \label{qmetric}
g_{\mu \nu}  = \R(\partial_\mu \psi, \partial_\nu \psi),
\end{equation}
where $\R$ indicates the real part, $||\cdot||$ indicates the norm defined by the scalar product, and $\mu, \nu$ run over $x$ and $t$, i.e. the spatial and temporal coordinates. For more details on quantum mechanical metrics, see \cite{Provost}.

Considering now a position vector on the soliton surface $\bold r_\lambda(x,t)$, where $\lambda$ is the spectral parameter of the IST, the metric (\ref{metric1}) can be recast in the form \cite{rogers}
\begin{equation} \label{cmetric}
g_{\mu \nu} = \R(\partial_\mu \bold r_\lambda, \partial_\nu \bold r_\lambda),
\end{equation}
where the real part is needed to describe a position on a real soliton surface since the spectral parameter might be complex (depending on the point spectrum of the direct problem of the IST) as in the case of NLSE \cite{Zakharov3}. Equations (\ref{qmetric}) and (\ref{cmetric}) clearly have the same structure. Roughly speaking the manifold of quantum states in a Hilbert space is represented in the case of soliton surfaces by the the manifold of the position vectors on the surface. This supports our claim that any fluctuations (either classical or quantum) propagating on the body of the soliton perceives a space-time metric given by Eq. (\ref{metric1}), and therefore the soliton surface metric is not just a mathematical construction, but a fully physical one. We shall see that this observation has far-reaching consequences in terms of our idea to identify solitons and black holes in two dimensions.

Having introduced the general form of the metric (\ref{metric1}) we are now in a position to calculate the curvature invariants by using the standard methods of Riemannian geometry.

In general the Riemann tensor has $d^{2}(d^{2}-1)/12$ independent components, which for a dimensionality $d=2$ reduces to one independent component. The first quantities to consider are of course the traditional Ricci tensor $R_{\mu\nu}\equiv g^{\rho\tau}R_{\mu\rho\nu\tau}$ (where $R_{\mu\rho\nu\tau}$ is the Riemann curvature tensor, see for instance Ref. \cite{chandra}) and the Ricci scalar $R\equiv g^{\mu\nu}R_{\mu\nu}$. In 2D, these quantities are always related by the equation $R_{\mu\nu}-g_{\mu\nu}R/2=0$, i.e. the Einstein equations are always satisfied in 2D, with no matter present -- classically the energy-momentum tensor $T_{\mu\nu}$ must vanish identically in 2D, due to the reduced number of independent components of the Riemann tensor, and thus the Einstein equations are classically trivial \cite{note}. Another scalar of importance is the Kretschmann scalar $K_{1}\equiv R^{\mu\nu\rho\tau}R_{\mu\nu\rho\tau}$, which is also not independent from the Ricci scalar in 2D, and in fact the equivalence $K_{1}=R^{2}$ always holds (in both Minkowski and Euclidean metrics). The Kretschmann scalar is used, in 4D Schwarzschild spacetime, to show that the horizon is not a physical singularity but a fictitious one, since this scalar does not diverge when traversing the horizon, but blows up at the centre of the black hole, where the real classical singularity is located \cite{chandra}.

A more important scalar for our purposes in this paper, the study of which is quite neglected in textbooks treating classical four-dimensional general relativity, is the so-called Karlhede scalar, $K_{2}\equiv R^{\mu\nu\rho\sigma;\tau}R_{\mu\nu\rho\sigma;\tau}$, where the symbol ``;" indicates a covariant derivative \cite{chandra,carroll}. The roots of the Karlhede scalar are known to provide a means to detect the positions of the event horizons of a spherically symmetric black hole (for instance a \schw or a \rn black hole), where $K_{2}$ vanishes and changes sign after traversing the horizons. However, for some types of black hole, $K_{2}$ is known to possess extra roots that are associated to other regions not related to any event horizons -- for example in the Kerr metric, which describes uncharged rotating black holes, $K_{2}$ vanishes at the so-called ergosphere \cite{karlhede_scal}. Care must be taken in this regard also when dealing with the metric for solitons, see below for a discussion.

As an example of the use of these scalars for black hole physics in 2D, if one chooses to use the Schwarzschild-like metric $ds^{2}=\pm f(r)dt^{2}+f(r)^{-1}dr^{2}$ (where the $+$ or $-$ refers to Euclidean or Minkowski metrics respectively), with $r$ and $t$ the two \schw coordinates and $f(r)$ the \schw function only depending on $r$, then the above scalars can be expressed in the following way: $R=-f''$, $K_{1}=(f'')^{2}$ and $K_{2}=f(f''')^{2}$, where the primes indicate derivatives with respect to $r$. Similar to the 4D case, the points $K_{2}=0$ indicate the positions of the event horizons (defined by the equation $f(r)=0$) but, due to the presence of $f'''$, one must be careful to exclude those points for which $f'''=0$, which do not represent horizons. For example, in the sine-Gordon soliton case $K_{2}$ is not useful as $f'''$ vanishes identically, while in the KdV soliton case $f'''$ does not vanish identically but has three real roots not associated to event horizons. From these considerations it is clear that the key ingredient is the \schw function $f$, the only meaningful degree of freedom in 2D -- nonetheless the use of $R$ and $K_{2}$ can often be useful and illuminating.

Although in most cases it is quite difficult to guess the coordinate transformations leading to a Schwarzschild-like metric, a general method applicable in 2D was discussed by Chandrasekhar \cite{chandra}. By solving the so-called {\em Laplace-Beltrami} equation, it is {\em always} possible in 2D to find a coordinate transformation leading to a conformal metric, from which it is easy to perform an ``inverse tortoise" coordinate transformation that puts it in a Schwarzschild-like form \cite{susskind}. Solving the Laplace-Beltrami equation may be feasible in some solitonic metrics, and quite difficult in others. For the cases of the NLSE, KdV and sine-Gordon systems studied in this work we directly show the right transformations leading to Schwarzschild-like coordinates by first diagonalising the metrics.

\section{The NLSE and Hawking radiation} \label{sec6}

The first important representative example for our method is the bright soliton solution of the NLSE, Eq. (\ref{nlse1}). The starting point is the knowledge of the AKNS matrices for this equation, Eqs. (\ref{akns1}-\ref{akns2}). The classical metric describing the line element of the Hashimoto surface for the NLSE is found by applying Eq. (\ref{metric1}) with $\alpha=-2$, since $\hat{g}_{1,2}$ are elements of $\mathfrak{su}$(2). Eq. (\ref{comp1}), the compatibility condition for these matrices, directly generates the NLSE, see Eq. (\ref{nlse1}).

A straightforward calculation based on Eq. (\ref{metric1}) shows that the metric is
\eq{metricNLSE}{g_{\mu\nu}=\left(\begin{array}{cc}
4 & -16\lambda \\
-16\lambda & 16\left( |u|^2+4\lambda^2 \right)
\end{array}\right),} where $u(x,t)$ is the bright soliton solution of Eq. (\ref{nlse1}), given by
\eq{solu}{u=B{\rm sech}\left[B\left(x - vt\right)\right]e^{i\left[ \frac{v}{2}x + \left( B^2- v^2/4 \right) t \right]},}
where $B$ is the soliton amplitude, $v$ its velocity (the parameter space for this soliton is therefore two-dimensional), and $\lambda$ is the spectral parameter which encodes the information on the soliton parameters $B$ and $v$. The line element corresponding to the metric (\ref{metricNLSE}) is: $ds^2=4dx^2 - 32\lambda dx dt + 16 \left( |u|^2 + 4\lambda^2 \right)dt^2$. Note that the soliton phase in Eq. (\ref{solu}) in unimportant in defining the metric since only $|u|^{2}$ appears in Eq. (\ref{metricNLSE}).

The Ricci scalar for this metric, with static $v=0$ soliton inputted, is given by $R=-B^{2}/2+B^{2}{\rm sech}^{2}(Bx)$, which is constant and negative for large values of the argument $|x|$ (i.e. on the soliton tails), while it is positive in the proximity of the soliton peak $|x|=0$. This can be seen in Fig. \ref{fig1}(a) where the soliton peak generates a doughnut ring around the Beltrami pseudosphere (a surface with a constant negative Gaussian curvature). For the moving soliton solution with velocity $v\neq 0$, to be introduced later in Section \ref{sec8}, the Hashimoto surface is more complex and looks more like ``fusilli pasta", see Fig. \ref{fig1}(b). Initially, $\lambda$ in Eq. (\ref{nlse1}) is assumed to be real, however the IST method imposes an analytical continuation on the complex plane in order to ensure the convergence of the so-called Jost functions in the scattering problem (see Ref. \cite{jianke}, pp. 17-32). Therefore $\lambda$ becomes complex, and must acquire the value $\lambda=-v/4+iB/2$, which is derived using the IST (see Ref. \cite{jianke}, page 34). However, after substituting this value into the metric Eq. (\ref{metricNLSE}), one must take the real part of the metric tensor in order to be able to describe a real Hashimoto surface in accordance with eq. (\ref{cmetric}), obtaining:
\eq{metricNLSE2}{g_{\mu\nu}=\left(\begin{array}{cc}
4 & 4v \\
4v & 16|u|^2+4v^{2}-16B^{2}
\end{array}\right).}

The ``injection"  of the spectral parameter $\lambda$ is a transformation into the moving frame of the soliton \cite{rogers} and gives the metric above a dependence on the soliton velocity and amplitude. Due to this velocity dependence, the metric describes the geometry of the soliton already in its moving frame. Therefore the soliton solution to be inputted in the metric (\ref{metricNLSE2}) is the stationary one: $u=B{\rm sech}(Bx)e^{iB^2 t}$.

Let us use the scalar invariants introduced in Section \ref{sec3} to study the metric (\ref{metricNLSE2}) in these comoving coordinates. The metric determinant is $g=-64B^{2}\tanh^{2}(Bx)<0$, which means that the space defined by the Hashimoto surface has a Minkowski signature. The Ricci scalar is $R=|u(x,t)|^{2}>0$ and the Kretschmann scalar is $K_{1}=R^{2}=|u(x,t)|^{4}$. Note that the background negative curvature of the Beltrami pseudosphere has disappeared when passing in the comoving frame after the spectral parameter injection. The Karlhede scalar, whose zeros determine the horizon positions, is given by $K_{2}=-(1/4)|u|^{4}[v^{2}+4(|u|^{2}-B^{2})]$, which vanishes at two points $x_{\rm H,\pm}=\pm B^{-1}{\rm arctanh}[\frac{v}{2B}]$ symmetrically placed around the soliton peak, giving the location of the horizon on the body of the soliton. The reason that there are two points yet we speak of only one horizon is that a horizon in one spatial dimension can be thought of as a circle in one spatial dimension, that is two disconnected points. Another way to think of these two points is that they are entrance points into the event horizon itself. Note that for $v=0$ the two points defining the horizon coalesce into one, located at the top of the soliton peak $x=0$. For a velocity $|v|<2B$, the two points split symmetrically around the soliton maximum, while for $|v|=2B$ the points shoot out to infinity, one at $x=+\infty$ and the other one at $x=-\infty$, on the soliton tails. This introduces a finite maximum velocity above which the horizon disappear. Note that the existence of this maximum velocity is related to the approximations used to derive the NLSE from a second-order wave equation, namely the slowly-varying envelope approximation (SVEA), and reaching a velocity $|v|=2B$ would violate the assumptions used in SVEA, namely the temporal phase in Eq. (\ref{solu}) would vanish and then change sign.

Let us now put (\ref{metricNLSE2}) in Schwarzschild-like form. First we simplify somewhat by defining $\rho \equiv B x$, i.e. changing coordinates $(x,t)\rightarrow(\rho,t)$, giving a line element $ds^2=4B^{-2}d\rho^2 + 8vB^{-1}dt d\rho + \left( 4v^2 -16B^2 {\rm tanh}^2(\rho) \right) dt^2$. Then performing the transformation $(\rho,t)\rightarrow(\rho,\tau)$, using
$\tau \equiv t + \{4B{\rm arctanh}[2v^{-1}B{\rm tanh}(\rho)]-2v\rho\}/[8B^3 - 2Bv^2]$,
leads to the diagonal metric: $ds^2= \left( 16{\rm tanh}^2 (\rho)[4B^2 {\rm tanh}^2 (\rho) - v^2]^{-1} \right) d\rho^2 + \left( 4v^2 - 16B^2 {\rm tanh}^2 (\rho) \right) d\tau^2$. Finally, we transform to Schwarzschild-like coordinates, $(\rho,\tau)\rightarrow(r,\tau)$, with $r\equiv 8\ln(\cosh(\rho))$, obtaining the line element $ds^{2}=-f(r)d\tau^{2}+f(r)^{-1}dr^{2}$, with $f(r)\equiv 16B^2\left( 1-e^{-r/4} \right) - 4v^2$.

\begin{figure}[h]
\includegraphics[width=9cm]{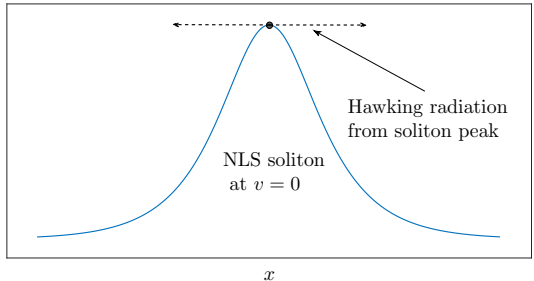}
\caption{A schematic of the NLSE soliton without velocity, showing its event horizon position at the soliton peak.
\label{fig2a}}
\end{figure}

\begin{figure}[h]
\includegraphics[width=9cm]{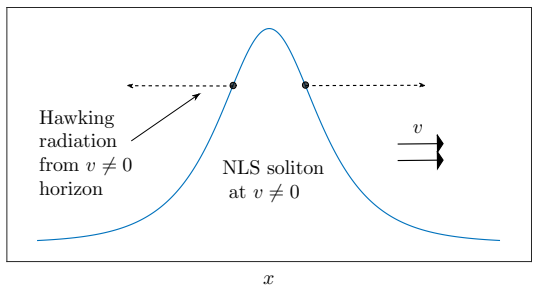}
\caption{A schematic of the NLSE soliton with velocity, showing its event horizon position located at two points placed symmetrically with respect to its peak. \label{fig2b}}
\end{figure}

The event horizon position $r_{\rm H}$ can be read directly off of the Schwarzschild-like metric, using $f(r_{\rm H})=0$, and is $r_{\rm H}=4{\rm ln}\left( 4B^2[4B^2-v^2]^{-1} \right)$. In terms of the coordinate $\rho$ defined above the horizon positions are $\rho_{\rm H,\pm}=\pm{\rm arctanh}[v/(2B)]$. Again, these diverge towards the soliton tails as $v\rightarrow \pm 2B$, indicating a speed limit above which the soliton cannot maintain any horizon. Approaching the stationary case, as $v\rightarrow 0$, the horizon closes and for $v=0$ is located at the peak of the soliton. The positions of the horizon at zero and non-zero velocities $v$ are sketched diagrammatically in Figs. \ref{fig2a} and \ref{fig2b}.

The Karlhede scalar defined in Section \ref{sec3} can be again evaluated and is, in the Schwarzschild-like coordinates $(r,\tau)$, equal to $K_{2}=(1/4)e^{-3r/4}\left( -4B^{6} + B^{4}e^{r/4}\left( 4B^2 - V^2 \right) \right)$. It vanishes at the event horizon position $r_{\rm H}=4{\rm ln}\left( 4B^2[4B^2-v^2]^{-1} \right)$, in exact agreement with the above calculations. The great advantage of using the Karlhede scalar for determining the horizon is that the calculation is {\em covariant}, and therefore one does not need to diagonalise the metric tensor to a Schwarzschild-like form, but  one can choose any coordinate system to perform the calculation, as we have shown above. The (slight) disadvantage is that covariant derivatives of the Riemann tensor must be evaluated, but this is readily done with any good symbolic software.

An interesting fact that distinguishes the black-hole structure of the NLSE from the conventional Schwarzschild black hole is that the former does not possess any singularity, while the latter contains an unremovable singularity at $r=0$.

\begin{figure}[h]
\includegraphics[width=9cm]{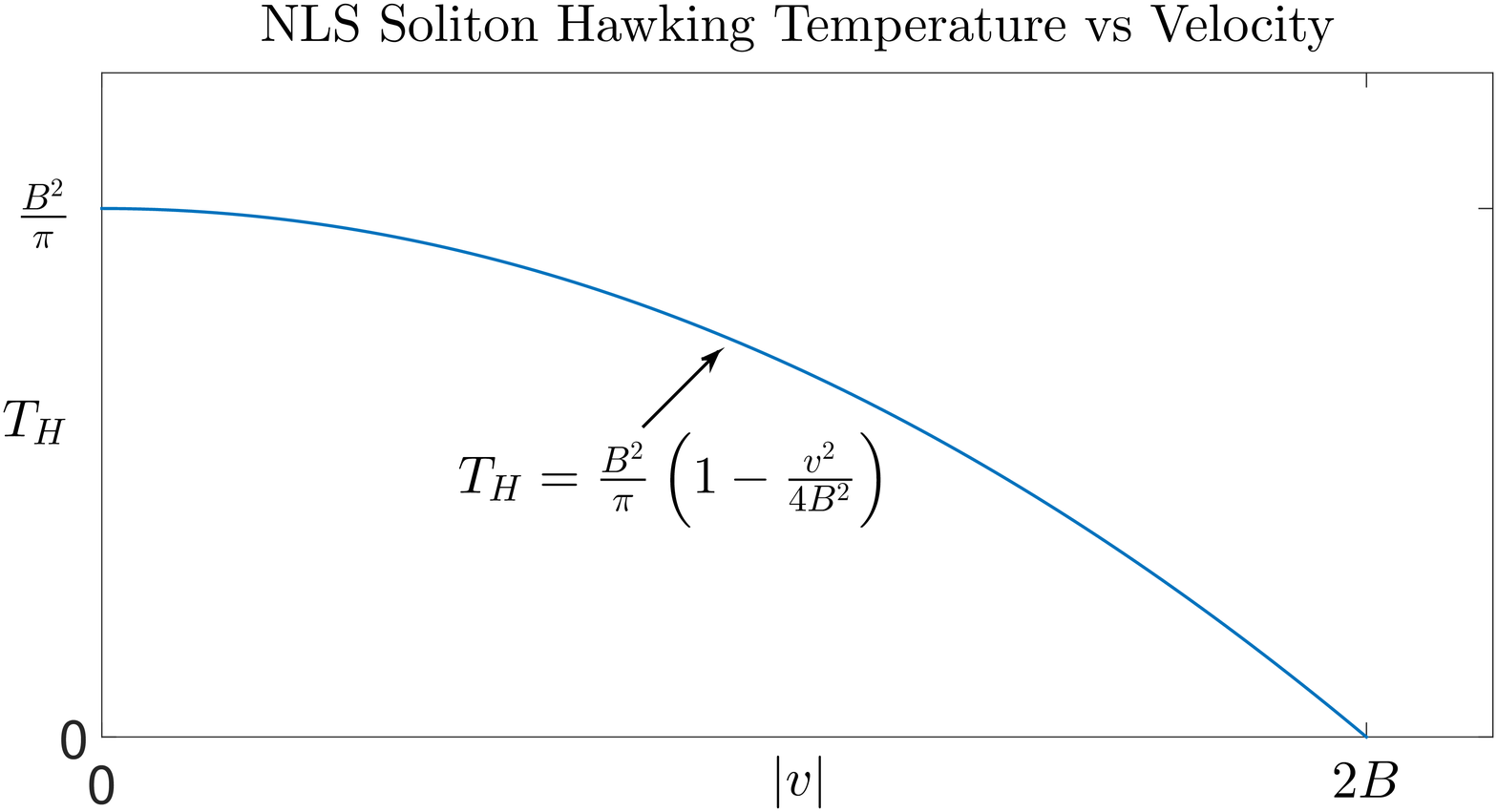}
\caption{NLSE soliton Hawking temperature as a function of its linear velocity. \label{fig3}}
\end{figure}

We now use the standard formulas for finding the Hawking temperature (in dimensionless units) in two dimensions, which can be verified using the anomaly methods used in Schwarzschild-like coordinates \cite{carroll,setare,tong}. The `surface gravity' is defined as $\Sigma\equiv |f'(r_{\rm H})|/2$ and the Hawking temperature is $T_{\rm H}=\Sigma/(2\pi)$. One finds the following NLSE soliton temperature:
\begin{equation} \label{eq:NLS_temp}
T_{\rm H}=\frac{B^2}{\pi}\left( 1 - \frac{v^2}{4B^2} \right).
\end{equation}
The soliton temperature form is plotted in Fig. \ref{fig3}, showing how it approaches zero when $|v| \rightarrow 2B$.

\section{The case of the KdV soliton and sine-Gordon kink} \label{sec7}

We now consider the case of the fundamental KdV soliton. We use the following representation of the AKNS matrices, based on the Lie algebra of $2\times 2$ real traceless matrices $\mathfrak{sl}_{2}(\mathbb{R})$ \cite{aktosun} (another representation based on the Lie algebra $\mathfrak{su}$(2) is also available, based on the traditional forms given in most references \cite{ablowitz}, but it is more difficult to use -- changing representation does not lead to a loss of generality):
\begin{eqnarray}
\hat{g}_{1}&=&\left(\begin{array}{cc} 0 & u-\lambda \\ 1 & 0 \end{array}\right), \label{aknskdv1} \\
\hat{g}_{2}&=&\left(\begin{array}{cc} u_{x} & -4\lambda^{2}+2\lambda u+2u^2 \\
4\lambda+2u & -u_{x}  \end{array}\right), \label{aknskdv2}
\end{eqnarray}
with the compatibility condition (\ref{comp1}) satisfied if $u_{t}-6uu_{x}+u_{xxx}=0$, which is the KdV equation, used to study amongst other things nonlinear water waves \cite{drazin}. Its real fundamental soliton solution is $u=-2B^{2}{\rm sech}^{2}(B(x-4B^{2}t))$, and therefore the parameter space for this soliton is 1-dimensional, given by the value of the amplitude $B$. The spectral parameter is purely real and imposed to be $\lambda=-B^{2}$ by the IST technique.

The line element associated to this solution is, using Eq. (\ref{metric1}), given by $ds^{2}=16dxdt-32(u-4\lambda)dt^{2}$. The soliton surface connected to this metric is shown in Fig. \ref{fig1}(c). As in the NLSE case above, the appropriate soliton to input into the metric is a stationary one. Note that the above KdV soliton solution cannot have its velocity set to zero arbitrarily as its velocity is tied to its amplitude. In that case, one needs to use the following stationary solution (taking the appropriate limit of Eq. (1.4) on page 2 in Ref. \cite{grimshaw}) $u(x)=-2B^{2}{\rm sech}^{2}(Bx) + 2B^{2}/3$, where a non-vanishing background appears in the comoving frame -- the background is necessary to ensure that $u(x)$ is still a solution of the time-independent KdV.
We will now prove that this metric possesses no horizon. If one uses the Karlhede scalar one finds three real roots, however these roots belong to the zeros of $f'''$, where $f$ is the \schw function for the KdV. Since the Karlhede scalar in \schw coordinates is given by $K_{2}=f(f''')^{2}$, one must exclude these roots in order to evaluate the real event horizon positions. In order to find the \schw function, we first use the coordinate transformation $(x,t)\rightarrow(\rho,t)$, with $\rho\equiv Bx$, and then apply the following transformations: $(\rho,t)\rightarrow(\rho,\tau)$, with $d\tau\equiv dt+3d\rho\left[8B^{3}(-7+3{\rm sech}^{2}(\rho))\right]^{-1}$, and $(\rho,\tau)\rightarrow(\rho,T)$, with $dT\equiv-4\sqrt{2}d\tau [1+7{\rm cosh}(2\rho)]^{1/2}/[B{\rm sech}(\rho){\rm cosh}(\rho)\{-3+7{\rm cosh}^{2}(\rho)\}^{1/2}]$ and find that in the $(\rho,T)$ coordinates the line element is $ds^{2}=-f(\rho)dT^{2}+f^{-1}(\rho)d\rho^{2}$, with $f(\rho)\equiv(B^{4}/3)(-3+7{\rm cosh}^{2}(\rho)){\rm sech}^{2}(\rho)$, which is in the desired form. Thus $f=0$ has no real solutions, meaning that there is no physical event horizon associated to the KdV fundamental soliton, and so it does not emit any Hawking radiation.

The sine-Gordon equation $u_{tt} - u_{xx} + \mathrm{sin}(u)=0$ is another integrable equation with a soliton solution, known as a kink, with known links to black hole theory in 2D \cite{gegenberg}. This equation has found uses in differential geometry, relativistic field theory and Josephson junction theory, amongst other places \cite{williams}. Its IST-induced geometry has been studied before by Gegenberg \cite{gegenberg} and later in great detail by one of the authors \cite{leone1} and was found to describe a 2D black hole with Hawking temperature $T_{\rm H}=v/(2\pi)$ where $v$ is the velocity of the kink. All of the relevant formulas, including AKNS matrices and coordinate transformations leading to a \schw form of the metric can be found in Ref. \cite{leone1}. We include this result here to contrast with our NLSE soliton Hawking temperature found in the previous section. For the NLSE soliton, a stationary soliton with $v=0$ still gives a non-zero temperature, whereas the sine-Gordon black hole temperature vanishes at $v=0$.

It is very important to observe at this point that the above procedure can be directly applied to `conventional' black holes of {\em any dimensionality} $\geq 2$. By choosing coordinates close to the (outer) event horizon and keeping only the dominant terms of the expansion of the action of a scalar field embedded in the gravitational field of the black hole, one can typically reduce the high-dimensional metric to a two-dimensional one (see for instance Ref. \cite{redux} for a particularly lucid discussion). This can be done for spherically symmetric and surprisingly even for non-spherically symmetric black holes, like the Kerr metric \cite{redux}, allowing the methods described here to be used.

In the next section we show another (more heuristic, but very effective) approach to the calculation of the Hawking temperature for solitons, as well as their entropies, based on the first law of thermodynamics applied to integrable equations.

\section{Quantum soliton thermodynamics} \label{sec8}

We now introduce for the first time the concept of {\em quantum soliton thermodynamics}, deriving the first law of thermodynamics (in differential form) associated to the quantum emission of Hawking radiation for the NLSE and KdV solitons as well as the sine-Gordon kink. This concept is well suited for solitons of integrable equations: in fact, for these class of equations, there is a strong correspondence between the conserved quantities of a mechanical particle (either classical or relativistic) and the conservation laws for solitons \cite{hasse}.

\subsection{NLSE soliton thermodynamics}

We begin with the most physically interesting case: the NLSE soliton. To acquire the thermodynamical first law for this soliton we begin by finding its Hamiltonian in terms of its parameters, amplitude $B$ and velocity $v$. The conserved quantity of the NLSE Eq. (\ref{nlse1}) associated with the Hamiltonian is known to be $H= \int_{-\infty}^{\infty}\left( -|u_{x}|^2 + |u|^4 \right)dx$ which for the moving soliton solution $u=B{\rm sech}(B(x - vt))e^{i\left( \frac{v}{2}x + \left( B^2- v^2/4 \right) t \right)}$ gives:
\begin{equation} \label{NLS_thermo_1}
H=\frac{2B^3}{3}-\frac{Bv^2}{2}.
\end{equation}

The differential of the Hamiltonian immediately gives the first law of thermodynamics for the moving soliton. One calculates the differential using the chain formula $dH=\frac{\partial H}{\partial B}dB + \frac{\partial H}{\partial v}dv$, giving:
\begin{equation} \label{eq:firstlaw1}
dH=\left( 2B^2 - \frac{v^2}{2} \right) dB - (Bv) dv,
\end{equation}
with the first term representing a heat term and the second a work term. Rearranging gives:
\begin{equation} \label{eq:firstlaw2}
dH=\frac{B^2}{\pi}\left( 1-\frac{v^2}{4B^2} \right) dS - dW,
\end{equation}
where $W\equiv Bv^{2}/2$ is the work done by the system. This fixes the soliton entropy to be $S=2\pi B$. Note that $W$ resembles the kinetic term of a mechanical particle, with $B$ playing the role of the particle mass. This is consistent with the particle-soliton analogy concept for solutions of integrable equations \cite{hasse}.
In fact, using two other known NLSE conservation laws, the soliton momentum is found to be $P\equiv i\int_{-\infty}^{+\infty}(u^{*}_{x}u-u_{x}u^{*})dx=Bv$, and its mass $M\equiv(1/2)\int_{-\infty}^{+\infty}|u|^{2}dx=B$ and so the work term differential above, $dW$, is equal to $Pdv$. The conservation law for mass shows that $B=M$, meaning that the soliton entropy is related to its mass by $S=2\pi M$. This is a known relation describing the entropy of 2D black holes -- described as the generalization of the Bekenstein-Hawking entropy in 2D \cite{giddings,cadoni} -- which we have re-derived here purely from studying the NLSE soliton thermodynamics. This once again emphasises the deep links between solitons and black holes. It can be seen that the heat term in (\ref{eq:firstlaw2}) is of the form $T_{\mathrm{H}}dS$ and remarkably this reproduces the Hawking temperature found from our NLSE-induced geometry method, Eq. (\ref{eq:NLS_temp}).

An interesting feature concerning the moving soliton energy: if one wishes to ensure positive soliton energy $H \geq0$, then rearranging the formula for $H$ given above shows that the following must always hold: $v^2 \leq (4/3)B^2$. Clearly this velocity limit is lower than the one required for radiation emission to occur, $v=2B$, and thus the soliton must have a temperature and radiate at all times.

\subsection{KdV soliton thermodynamics}

The conservation laws for the KdV soliton (see Ref. \cite{drazin} for full details and definitions) define a Hamiltonian $H\equiv -5\int_{-\infty}^{+\infty}[u^{3}+(1/2)u_{x}^{2}]dx=32B^5$, momentum $P\equiv 3\int_{-\infty}^{+\infty}u^{2}dx=16B^3$, and mass $M\equiv-\int_{-\infty}^{+\infty}udx=4B$, while the soliton velocity can be read off the form of the solution $u=-2B^{2}{\rm sech}^{2}(B(x-4B^{2}t))$ and is $v=4B^2$. Note that $H=Mv^{2}/2$ as for a classical point particle. Therefore:
\begin{equation}
dH=\frac{dH}{dB}dB=160B^{4}dB,
\end{equation}
which can be interpreted as a pure work term. This work term leaves no room for a further heat term which, along with the KdV metric results presented in Section \ref{sec7}, evinces that no Hawking radiation emission occurs for the KdV soliton, and therefore there is no Hawking temperature due to the absence of an event horizon.

\subsection{Sine-Gordon kink thermodynamics}

Following the above procedure, the sine-Gordon equation has a first law of thermodynamics (see Ref. \cite{dutykh} for details on the conserved quantities of the sine-Gordon):
\begin{equation} \label{eq:sG_first_law}
dH=\frac{dH}{dv}dv=vdP
\end{equation}
where Hamiltonian $H=8/\sqrt{1-v^2}$ and momentum $P=8v/\sqrt{1-v^2}$, which are energy and momentum of a relativistic particle of mass $M=8$ in terms of a unit `speed of light'. The right hand side of (\ref{eq:sG_first_law}) should be interpreted as a heat term in line with the known result that the sine-Gordon kink radiates at temperature $T_{H}=v/(2\pi)$ \cite{leone1}. This first law thus gives an entropy linked to momentum as $dS=2\pi dP$ or $S=2\pi P$, again showing the $2\pi$ dependence of the entropy on a soliton parameter in two dimensions.

Interpreting which term of each first law represents heat and which term represents work is subtle and should be considered on a case-by-case method, taking into account the role of the soliton parameters for each integrable system, and combining the thermodynamical arguments with evidence from their induced metrics.

The thermodynamical methods introduced here are a powerful tool for finding features of soliton solutions such as temperature and entropy which in many cases may be very difficult to determine otherwise. We have shown that the soliton conserved quantities (such as energy, momentum and mass), in terms of its parameters, combined with the soliton/particle correspondence principle valid for integrable equations, contain all of the information needed for this analysis.

\section{Temperature, entropy and lifetime of the NLSE soliton in dimensional units} \label{sec9}

\subsection{Metric, temperature and entropy in dimensional units} \label{sec9a}

In this section we perform the previous analysis of the properties of the NLSE soliton in dimensional units, finding the Hawking temperature and entropy for a real soliton that can be studied in the laboratory. The use of dimensional units considerably clarifies the physics of the quantum evaporation of solitons. We specialise our discussion to a fiber optics soliton, giving detailed numerical estimates of its Hawking temperature and lifetime.

The NLSE with units introduced is given by $i\psi_{t}+\omega '' \psi_{\xi \xi} - 2\kappa |\psi|^2 \psi=0$ (with $\kappa<0$) and its associated line element is:
\begin{equation} \label{eq:units_ln_elem}
\begin{split}
ds^2 = & 4\widetilde{\kappa}^2 k_{0}^2 d\xi^2 - 32 \widetilde{\kappa}^3 \lambda ' k_{0}^3 \omega '' d\xi dt \\
& - 2\widetilde{\kappa}^2 \left( k_{0}^2 \omega '' \right)^2 \left( -8\widetilde{\kappa}|\psi|^2 -32\widetilde{\kappa}^2 {\lambda ' }^2 \right) dt^2,
\end{split}
\end{equation}
where $\xi$ and $t$ are dimensional space and time variables. The frequency dispersion is $\omega(k)$, the wave vector and central wave vector are $k$ and $k_{0}$, dispersion parameter $\omega''=\frac{\partial^2 \omega(k)}{\partial k^2}\Big|_{k=k_0}$, $\widetilde{\kappa}=|\kappa|/(\omega'' k_{0}^2)$, and the soliton spectral parameter (in dimensional units) is defined as $\lambda' = \sqrt{\frac{2}{|\kappa| \widetilde{\kappa}}}( -\phi + i\psi_{0} )$, where $\phi=v_{s}/(4\sqrt{2\omega''})$, $\psi_{0}=\frac{|\kappa|}{8}\sqrt{\frac{2}{\omega''}}N$, $v_{s}$ is a velocity parameter (whose meaning differs for spatial and temporal solitons) and $N$ is average photon number in the soliton. The nonlinear parameter $\kappa$ is defined as $\kappa=-\frac{3\hbar \chi^{(3)} \omega_{0}^2 v_{g}^2}{4\epsilon \mathcal{A}c^2}$ (its role in the NLSE Hamiltonian is shown below), where $\chi^{(3)}$ is the third-order nonlinear susceptibility of the material supporting the soliton, $v_{g}$ is the group velocity, $\mathcal{A}$ is the effective mode area, $\epsilon$ is the permittivity of the medium, $c$ is the speed of light in vacuum, and $\omega(k_{0})=\omega_{0}$.

The soliton solution itself in units is given by:
\begin{widetext}
\begin{equation} \label{eq:sol_w_units}
\psi (\xi,t) = \frac{2\sqrt{2}\psi_{0}}{\sqrt{|\kappa|}}e^{-8i(\phi^2 - \psi_{0}^2)t + 2i\phi \sqrt{2/\omega''}\xi}{\rm sech}\left( 2\psi_{0} \left( \sqrt{\frac{2}{\omega''}}\xi -8\phi t \right) \right).
\end{equation}
\end{widetext}
In an experiment with a spatial soliton, the velocity parameter $v_{s}$, the transverse drift of the soliton with respect to its origin position, will be small and can be taken as zero as for a non-drifting soliton. Substituting the soliton solution (\ref{eq:sol_w_units}) into the line element (\ref{eq:units_ln_elem}), and defining a new variable $T=ct$, where $c$ is the speed of light, then taking its real part gives a diagonal form:
\begin{equation}
ds^2=4\widetilde{\kappa}^2 k_{0}^2 d\xi^2 - \frac{128\widetilde{\kappa}^3 \psi_{0}^2 k_{0}^4 ({\omega''})^2 {\rm tanh}^2 (\frac{2\sqrt{2}\psi_{0} \xi}{\sqrt{\omega ''}})}{c^2 |\kappa|}dT^2.
\end{equation}
The metric can now be put into \schw form and the Hawking temperature calculated this way, as we did in Section \ref{sec6}, however for a diagonal metric the following useful formula can be immediately used to calculate the surface gravity \cite{wu}, with units introduced:
\begin{equation}
\Sigma = -\frac{c^2}{2}\sqrt{\frac{g^{\xi \xi}}{-g_{TT}}}\frac{\partial g_{TT}}{\partial \xi}\Big|_{\xi=\xi_{H}}
\end{equation}
along with $T_{H}=\frac{\hbar \Sigma}{2\pi k_{B}c}$. This gives the Hawking temperature, in Kelvin, for the soliton:
\begin{equation} \label{eq:TH_units}
T_{H}=\frac{\hbar |\kappa|^2 N^2}{4 \pi \omega'' k_{B}}.
\end{equation}

A thermodynamical argument, introduced earlier without dimensions, verifies the validity of the above formula for the Hawking temperature as follows. The first law of thermodynamics with units is found by inserting soliton solution (\ref{eq:sol_w_units}) with $v_{s}=0$ into the Hamiltonian describing the system \cite{drummond}:
\begin{equation}
H = -\hbar \omega '' \int |\psi_{\xi}|^2  d\xi + \hbar |\kappa| \int |\psi|^4 d\xi,
\end{equation}
where all integrals are evaluated over all space. For the NLSE soliton one obtains:
\begin{equation}
H=\frac{32\sqrt{2 \omega''}\hbar \psi_{0}^3}{3 |\kappa|}.
\end{equation}
Substituting in our definition $\psi_{0}=\frac{|\kappa|}{8}\sqrt{\frac{2}{\omega''}}N$ produces: $H=\frac{N^3 \hbar |\kappa|^2}{12\omega''}$. As we showed in the dimensionless case in Section \ref{sec8} the natural route to a thermodynamical first law for the soliton is to find the differential of the Hamiltonian, which here can be cast as $dH=\frac{dH}{dN}dN$:
\begin{equation}
dH=\left( \frac{N^2 \hbar |\kappa|^2}{4\omega''} \right) dN.
\end{equation}
Comparing this with $dH=T_{H}dS$ for the Hawking temperature we have found (\ref{eq:TH_units}), verifies the form of our Hawking temperature and immediately gives the entropy differential as $dS=k_{B}\pi dN$, leading to the NLSE soliton entropy:
\begin{equation}
S=k_{B}N\pi.
\end{equation}
Not surprisingly, the total entropy only depends on the total average number of photons contained in the soliton, which are the only degrees of freedom in the system.

\subsection{The lifetime of the NLSE soliton} \label{sec9b}

An experimentally important question is: over which time and length scales does the NLSE soliton decay due to its quantum Hawking emission?
Following an analogous method used for conventional black holes, firstly we must derive an equation for the power of the emission, a Stefan-Boltzmann law in two dimensions. The simplest method is to use the results of Ref. \cite{landsberg}, where the Stefan-Boltzmann constants are given in general for any dimensionality. Using the same notation as in Ref. \cite{landsberg}, if $n$ indicates the number of spatial dimensions ($n=d-1$, where $d$ is the total number of dimensions), then the total power emitted by a blackbody is given by $P=A_{n}\sigma_{n}T_{\rm H}^{n+1}$, where $A_{n}$ is the $n$-dimensional black hole `area' and $\sigma_{n}$ is the Stefan-Boltzmann constant for $n$ spatial dimensions. Specifically one has $A_{1}=1$, $A_{2}=2\pi r_{\rm H}$ and $A_{3}=4\pi r_{\rm H}^{2}$, etc., while $\sigma_{1}=\pi k_{\rm B}^{2}/(6\hbar)$, $\sigma_{2}=2\zeta(3)k_{\rm B}^{3}/(\pi^{2}\hbar^{2}c)$ (where $\zeta$ is the Riemann-zeta function) and $\sigma_{3}=\pi^{2}k_{\rm B}^{4}/(60\hbar^{3}c^{2})$, etc., see Ref. \cite{landsberg}. The constants $\sigma_{n}$ are expressed in units of [Wm$^{n-1}$$^\circ$K$^{n+1}$]. Using $n=3$ one obtains the well-known Hawking evaporation lifetime of the \schw black hole, which shows that the black hole completely evaporates in a finite time \cite{lifetimeblackhole}.

Applying the above formulas for the 2D case, i.e. $n=1$, we obtain:
\begin{equation} \label{eq:power}
P=\frac{\pi k_{\mathrm{B}}^2 T_{\mathrm{H}}^2}{12\hbar}.
\end{equation}
The NLSE soliton loses energy $E$ as it radiates and so $P=-dE/dt$.

The nonlinear parameter $\kappa$, defined in Section \ref{sec9}, can be rearranged in the following way: using the standard nonlinear-optics definitions (see Ref. \cite{agrawal}) $\gamma\equiv n_{2}\omega_{0}/(\mathcal{A}c)$, $n_{2}\equiv(3\chi^{(3)}/4)[\epsilon_{0}cn^2]^{-1}$ and $\epsilon\equiv\epsilon_{0}n^2$, one finds $\kappa=-\hbar \omega_{0} \gamma v_{g}^2=-\hbar \omega_{0} \gamma/\beta_{1}^2$. Here $\gamma$ is the fiber nonlinearity, $n_2$ the nonlinear refractive index, $n$ is the linear refractive index, $\eps_{0}$ is the vacuum permittivity constant and $\beta_{1}$ is the inverse group velocity coefficient of the fiber. Defining the energy of the optical pulse to be $E(t)=\hbar \omega_{0} N(t)$ and using the NLSE soliton Hawking temperature found earlier (\ref{eq:TH_units}) we find:
\begin{equation}
T_{\mathrm{H}}=\frac{\hbar \gamma^2 E^2(t)}{4\pi k_{\mathrm{B}} \beta_{1}|\beta_{2}|}.
\label{hawkingfiber}
\end{equation}
Substituting this into the form of the radiative power (\ref{eq:power}), gives the energy loss as:
\begin{equation} \label{eq:energy_loss}
\frac{dE}{dt}=-\frac{\hbar \gamma^4}{192\pi \beta_{1}^2 |\beta_{2}|^2}E^4(t),
\end{equation}
or $dE/dt=-aE^4(t)$, where $a\equiv \hbar \gamma^4/[192\pi \beta_{1}^2 |\beta_{2}|^2]$. The energy and photon number at the initial time are $E_{0}=E(t=0)$ and $N_{0}=N(t=0)$.

\begin{figure}[h]
\includegraphics[width=9cm]{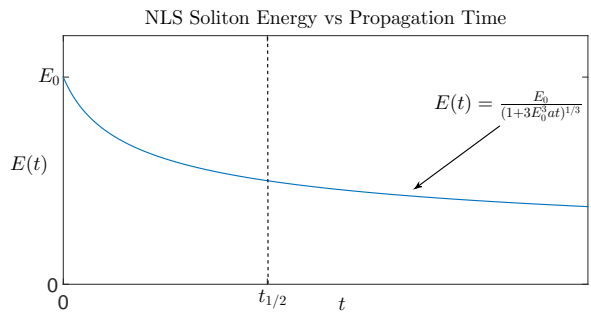}
\caption{The energy decay of the NLSE soliton due to Hawking radiation emission as a function of its propagation time. Its half-life is denoted by $t_{1/2}$. \label{fig4}}
\end{figure}

Solving (\ref{eq:energy_loss}) for the soliton energy $E$ as a function of time, one finds:
\begin{equation}
E(t)=\frac{E_{0}}{\left( 1+3E_{0}^3 a t \right)^{1/3}},
\end{equation}
plotted in Fig. \ref{fig4}.

Note that the soliton radiates by the Hawking process continuously, following an asymptotic decay law, in contrast with the well-known decay of \schw black holes where total decay occurs at a finite time \cite{lifetimeblackhole}. The precise way a soliton black-hole evaporates away obviously depends on the specific metric and solution considered. The ``half-life" of the NLSE soliton, that is the time it takes for its energy to halve via emission, is found easily from the form of $E(t)$ and is
\begin{equation}
t_{1/2}=\frac{64\pi \beta_{1}^2 |\beta_{2}|^2}{\hbar \gamma^4 (\hbar \omega_{0} N_{0})^3}.
\label{life1}
\end{equation}

The initial energy of the soliton obeys the well-known relation $E_{0}=2P_{0}t_{0}$ where $t_{0}$ is its width parameter and $P_{0}$ its initial peak power. Using these definitions along with $[\gamma P_0]^{-1}=t_{0}^2/|\beta_2|$ \cite{agrawal} one finds a half-life propagation distance of:
\begin{equation}
L_{1/2}=\frac{t_{1/2}}{\beta_1}=\frac{8\pi \beta_{1}t_{0}^3}{\hbar \gamma |\beta_{2}|}.
\label{life2}
\end{equation}

It is clear from Eq. (\ref{life2}) that in order to have any chance to observe quantum effects related to Hawking radiation in fibers one needs a high group-velocity dispersion parameter $\beta_{2}$, a small inverse group velocity parameter $\beta_{1}$, a short pulse duration $t_{0}$ and a fiber nonlinear coefficient $\gamma$ as large as possible. As an example, let us take a fiber soliton with a central wavelength $\lambda_{0}=1.55$ $\mu$m, a temporal duration of $t_{0}\simeq10$ fs, a highly nonlinear fiber with $\gamma\simeq 1$ W$^{-1}$m$^{-1}$, anomalous dispersion coefficient $\beta_{2}\simeq -10^{-24}$ s$^{2}$/m, $\beta_{1}\simeq 5\cdot10^{-8}$ s/m. It follows that the soliton peak power must be $P_{0}=|\beta_{2}|/(t_{0}^{2}\gamma)\simeq 10$ kW, associated to an average photon number $N\simeq 10^{9}$, and using Eq. (\ref{life2}) we obtain $L_{1/2}\simeq 10^{6}$ km. Even though this extremely long decay distance seems to be impossible to achieve with normal means, researchers have recently demonstrated that solitons in fiber cavities can travel for literally astronomical distances \cite{erkintalo}. Even though such solitons are not integrable and they are subject to gain and losses, we believe that at present cavity solitons represent the best chance to observe quantum effects related to the weak Hawking emission (a sort of `black hole in a cavity'). The Hawking temperature for the above parameters is calculated from Eq. (\ref{hawkingfiber}), and it is found to be $T_{\rm H}\simeq 5$ $^{\circ}$K, which is in any case presumably many orders of magnitude larger than any temperature of a real gravitational black hole observed until now.

\section{Conclusions}\label{conclusions}

In this paper, following Salam's original idea, we have shown that there is a deep and complete mathematical and physical analogy between black holes and solitons of integrable equations. Starting from the AKNS matrices of the equation and a specific localised soliton solution, one can construct a metric that defines a curved soliton surface that is perceived by the classical and quantum fluctuations propagating inside the soliton. This allows a general relativistic analysis of the curvature tensors and scalars associated with the soliton. When quantum effects come into play via the conformal anomaly, a Hawking temperature can be calculated for the horizon, leading to the concept of soliton entropy and quantum soliton thermodynamics. We have explained this procedure by examining three examples of solitons for the NLS, KdV and sine-Gordon equations. A specific dimensional calculation has been carried out for the NLSE soliton, providing formulas for the Hawking temperature, entropy and lifetime of the soliton under Hawking evaporation. For typical fiber parameters the length scale at which the evaporation occurs is very long, however using fiber cavity solitons propagating over astronomical distances could give some hope to observing phenomena related to the Hawking radiation of an NLSE-like soliton. Our theory opens up new venues for the investigation of quantum effects in soliton physics, and can easily be extended to a large class of real black holes of any dimensionality, since choosing a coordinate system near the event horizon typically allows a dimensional reduction to two dimensions.

\section*{Acknowledgements}
We would like to thank Prof. Claudio Conti and Giulia Marcucci (Rome), Prof. Ewan M. Wright (Tucson), Prof. Rina Kanamoto (Tokyo) and Prof. Daniele Faccio (Glasgow) for useful early discussions. This research was funded by the German Max Planck Society for the Advancement of Science (MPG). LDMV acknowledges support from EPSRC through CM-CDT.


\begin{thebibliography}{99}
\bibitem{note} In fact, a stronger statement can be made. Since the classical equations of gravity are trivial in 2D, and thus the Einstein-Hilbert action does not lead to a meaningful dynamics in this reduced dimensionality, one is forced to introduce an extra auxiliary field in the action, the so-called dilaton. It is possible to prove that every nonlinear evolution equation expressible in terms of AKNS matrices must be described by a 2D dilaton gravity action. This extremely important claim has a number of applications which are outside the scope of the present paper, and will be discussed in a future publication.
\end{thebibliography}
\end{document}